\documentclass{rspublic}

\input epsf
\usepackage{graphicx}
\usepackage{epsfig}
\usepackage{psfrag}
\usepackage{mathrsfs}
\usepackage{amsfonts}

\renewcommand{\baselinestretch}{2}

\begin{document}

\title[Adhesion Problems with Elastic Plates]{Continuum Elasticity Theory Approach for Spontaneous Bending and Twisting of Ribbons Induced by Mechanical Anisotropy}

\author[Z. Chen, C. Majidi, D.J. Srolovitz \& M.P. Haataja]{Zi Chen\thanks{To whom
    correspondence should be addressed.  E-mail:
    chen.z@seas.wustl.edu.}$^{1-3}$, C. Majidi$^2$, D.J. Srolovitz$^{5,6}$, M.P. Haataja$^{1,2,7,8}$}
    \affiliation{$^1$Department of Mechanical and Aerospace Engineering, Princeton University, Princeton, NJ 08544, USA\\
    $^2$Princeton Institute for the Science \& Technology of Materials (PRISM) \\Princeton University, Princeton, NJ  08544, USA\\
    $^3$Department of Biomedical Engineering, Washington University, St. Louis, MO 63130, USA\\
    $^4$Department of Mechanical Engineering, Carnegie Mellon University,  Pittsburgh, PA 15213, USA\\
    $^5$Institute of High Performance Computing, 1 Fusionopolis Way, Singapore 138632\\
    $^6$Department of Materials Science and Engineering, University of Pennsylvania, Philadelphia, PA 19104, USA\\
    $^7$The Program in Applied and Computational Mathematics, Princeton University, Princeton, NJ 08544, USA\\
    $^8$School of Mathematics, Institute for Advanced Study, Princeton, NJ 08540, USA
    }

\label{firstpage}

\maketitle

\begin{abstract}{Helices, Elasticity theory, Surface stress, Nanohelices, Tubules} Helical ribbons arise in many biological and engineered systems, often driven by anisotropic surface stress, residual strain, and geometric or elastic mismatch between layers of a laminated composite. A full mathematical analysis is developed to  analytically predict the equilibrium deformed helical shape of an initially flat, straight ribbon, with prescribed magnitudes and orientations of the principal curvatures when subjected to arbitrary surface stress and/or internal residual strain distribution.  The helix angle, radius, axis and chirality of the deformed helical ribbons are predicted with a comprehensive, three-dimensional model that incorporates elasticity, differential geometry, and variational principles.  In general, the mechanical anisotropy (e.g., in surface/external stress, residual strain or elastic modulus) will lead to spontaneous, three-dimensional helical deformations.  Ring shapes are formed when the principle axes of deformation coincide with the geometric axes of the ribbon.  The transition from cylindrical helical ribbons to purely twisted ribbons, or tubular structures is controlled by tuning relevant geometric parameters. Analytic, closed-form predictions of the ribbon shapes are validated with simple, table-top experiments.  This theoretical approach represent a tool to inform the design of materials and systems in order to achieve desired helical geometries on demand.
\end{abstract}

\renewcommand{\baselinestretch}{2}

\newcommand{\dx}{\mbox{${\bf d}_x$}}
\newcommand{\dy}{\mbox{${\bf d}_y$}}
\newcommand{\dz}{\mbox{${\bf d}_z$}}
\newcommand{\rx}{\mbox{${\bf r}_1$}}
\newcommand{\ry}{\mbox{${\bf r}_2$}}
\newcommand{\bN}{\mbox{${\bf N}$}}
\newcommand{\bP}{\mbox{${\bf P}$}}
\newcommand{\Ex}{\mbox{${\bf E}_x$}}
\newcommand{\Ey}{\mbox{${\bf E}_y$}}
\newcommand{\Ez}{\mbox{${\bf E}_z$}}
\newcommand{\dux}{\mbox{${\bf u}_1$}}
\newcommand{\duy}{\mbox{${\bf u}_2$}}
\newcommand{\drx}{\mbox{${\bf r}_1$}}
\newcommand{\dry}{\mbox{${\bf r}_2$}}
\newcommand{\strain}{\mbox{\boldmath $\gamma$}}
\newcommand{\andsp}{\mbox{$\quad\textrm{and}\quad$}}
\newcommand{\pfrac}[2]{\frac{\partial #1}{\partial #2}}
\newcommand{\Dfrac}[2]{\frac{\mathrm{d} #1}{\mathrm{d} #2}}

\section{Introduction}

Helical structures arise in a variety of natural systems, such as plant tendrils (Chouaieb et al. 2006; Gerbode et al. 2012)
, chiral seed pods (Armon et al. 2011; Forterre \& Dumais 2011)
, stork's bill awn (Abraham et al. 2011)
and twisted guts (Savin et al. 2011; Wyczalkowski et al. 2012)
, as well as nanoengineered polymer lamellae (Wang et al. 2011)
, twist-nematic-elastomer films (Sawa et al. 2011)
, ZnO nanohelices (Kong et al. 2003; Majidi et al. 2010)
, graphene nanoribbons and nanotubes (Kit et al. 2012; Cranford and Buehler 2011)
. They have since attracted scientific attention for their potential role in nanoelectromechanical systems (NEMS) (Bunch et al. 2007),
drug delivery and biological/chemical sensing(Cui et al. 2001; Smith et al. 2001;Zastavker et al. 1999),
magnetic field detection (Huang et al. 2005),
optoelectronics (Hwang et al. 2008),
and microrobotics (Abbott et al. 2009).
With recent advancements in nanotechnology, physicists and engineers can now grow helical nanoribbons (Kong and Wang 2003)
through a ``bottom-up'' approach  and have also begun exploring ways to fabricate helical nanoribbons of controllable morphology (Bell et al. 2007; Prinz et al. 2000; Zhang et al. 2005)
in a ``top-down'' manner (Cho et al. 2006).
Motivated by potential applications in proteomics and biological surface engineering, scientists have also explored methods to control the morphology of helical nanoribbons through hierarchical self-assembly of amphiphilic molecules (Zhang et al. 2002).
Helical nanoribbons may also play a key role in emerging technologies like programmable matter, which adopt a broad range of geometries and functionalities in a controlled and reversible manner.

In this manuscript, we examine the mechanics and morphology of thin, effectively two-dimensional elastic helical ribbons embedded in $\Re^3$ within a theoretical framework that accounts for elastic bending along both the length and width directions (Chen et al. 2011)
.
Previous efforts have largely focused on one-dimensional representations for use in molecular dynamics simulations (Lee et al. 2011; V¨¢nai and Zakrzewska 2004)
and  continuum calculations (Chouaieb et al. 2006; Healey and Mehta 2005,
Helfrich 1986; Fuhrhop and Helfrich 1993; Kirchhoff 1859)
that treat the helix as a space curve.  While this approach may be appropriate for systems like DNA (Bryant et al. 2003; Snir and Kamien 2005)
,
$\alpha$-helices (Shepherd et al. 2009)
,
and carbon nanocoils (Bandaru et al. 2007)
,
there exist an important class of unique helical morphologies in nature and engineering (Chung et al. 1993; Giomi and Mahadevan 2010; Fuhrhop and Helfrich 1993;
Lugomer and Fukumoto 2010; Selinger et al. 2004; Srivastava et al. 2010) 
that can only be captured with a two-dimensional representation that treats the helix as a ribbon of finite width.  The helical ribbon model, presented here, represents a two-dimensional generalization of the classical Stoney/Timoshenko model used to study planar bending (Stoney 1909; Suo et al. 1999; Timoshenko 1925; Zang et al. 2007).

In many micro- and nano-scale systems, the mechanics of helices is governed by surface stress and other residual stresses that are comparable to the elastic restoring forces of bulk deformation (Gurtin and Murdoch 1975; He and Lilley 2008; Li et al. 2010;
Lotz and Cheng 2005; Miller and Shenoy 2000; Lachut and Sader 2007) 
.
These stresses may arise from surface defects or imbalances in atomic coordination number, adsorbates, and/or surface reconstruction (Yi and Duan 2009).
The shape of the helix may be controlled not only by the magnitude of the surface stress but also the anisotropy of both the residual stress field and ribbon elasticity (Chen et al. 2011; Lotz and Cheng 2005; Wang et al. 2008; Ye et al. 2009;
Zhang et al. 2005; Zhang et al. 2006; Wang et al. 2011; Wang et al. 2012)
For example, helicity may be introduced by the orientation of the in-plane principle components of either the elastic stiffness or surface stress, or a combination of both.  The equilibrium shape may, in turn, be influenced by interactions with  gas phase composition (Bandaru et al. 2007)
, surface chemistry (Randhawa et al. 2010)
, temperature (Timoshenko 1925)
, solution pH (Zhang et al. 2010)
, or other environmental factors that might affect surface stresses.  For ribbon laminates composed of two or more layers, helicity can also be controlled by lattice mismatch or pre-stretching individual layers (Zhang et al. 2005)
.  Although the physical origins of these stress asymmetries are different, we will demonstrate that the mechanics of each system represents special cases of a more general mathematical description.

Typically, ribbon helicity arises because at least one of the mechanical elements (e.g., surface stress (Wang et al. 2008)
, external stress (Pokroy et al. 2009)
, residual strain, or elastic modulus (Zhang et al. 2005)
is anisotropic and the principle axes of curvature do not coincide with the principle geometric axes (length and width) of the ribbon.  For quaternary sterol solutions such as model bile, anisotropic stress may be caused by a mismatch in molecular packing between constituent layers (Zastavker et al. 1999)
, while for nano-engineered helices, the deformation may be driven by epitaxial strains (Hwang et al. 2008)
.   Note that helical deformations represent large deflections of thin plates and is therefore a complex, nonlinear geometric problem that is usually difficult to solve even in very simple cases (Landau and Lifshitz 1986)
.  However, there has been increasing interest in studying morphologically distinct deformations such as purely twisted ribbons (with saddle curvature) (Liu et al. 2009; Oda et al. 1999)
, cylindrical helical shapes (Zhang et al. 2005)
and ring configurations (Cho et al. 2010; Mei et al. 2009)
, as well as in the connections between them (Bellesia et al. 2008; Ghafouri and Bruinsma 2005;
Igli$\acute{c}$ et al. 2005; Selinger et al. 2001; Selinger et al. 2004)
. Although traditionally each geometrically distinct object (Igli$\acute{c}$ et al. 2005; Selinger et al. 2001)
has been studied separately, recent studies (Bellesia et al. 2008;
Ghafouri and Bruinsma 2005; Selinger et al. 2004)
have explored the transitions between purely twisted ribbons and cylindrical helical shapes.  In the present work, we show that all of the above-mentioned shapes, in fact, belong to the same geometric class, governed by only three tunable geometric parameters.  We demonstrate that  transitions between these different shapes occurs smoothly as the geometric parameters are varied.

In this paper, we first employ concepts from differential geometry to establish a mathematical description of a ribbon deformed with a given set of principle curvatures (magnitudes and orientations), and show that the centerline deforms into a circular helix with constant radius and pitch. The as-deformed ribbon morphology represents a class of two-dimensional surfaces controlled by three independent geometric parameters, which, when properly tuned, give rise to different shapes such as cylindrical helical ribbons, purely twisted ribbons and ring shapes.  Then, we employ continuum elasticity theory and stationarity principles to quantitatively link principle curvatures and surface stresses or residual strains, providing a capability to predict the morphology of stress/strain-induced helical ribbons.  Closed-form analytic predictions are obtained and validated with simple, table-top experiments where layers of pre-stretched elastic sheets are bonded together to form laminated ribbons.  Good agreement is achieved between our theoretical analysis, which uniquely predicts shapes that cannot be explained with classical rod or plate theories, and experiments with no adjustable parameters.  Also, because the ribbon bends in two directions with non-zero Gauss curvature, our analysis is beyond the scope of the classical Stoney formulation of planar bending of ribbons under surface stress (Stoney 1909)
. This paper provides detailed derivations of the geometrical and continuum elasticity model along with explicit results and discussions, and hence greatly expands on the work in a very brief companion manuscript on this topic published previously by the authors (Chen et al. 2011)
.

\begin{figure}[t]
\begin{center}
\includegraphics[width=5in]{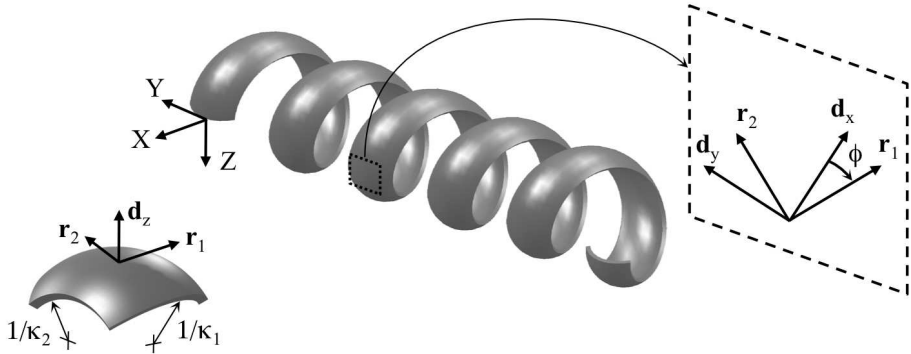}
\end{center}
\caption{Illustration of a helical ribbon.  The vectors $\dx$ and $\dy$ are oriented along the length- and width-wise axes of the ribbon, respectively.  The bases ${\bf r}_1$ and ${\bf r}_2$ correspond to the principle directions of curvature. }
\label{fig:helix}
\end{figure}

\section{Geometry of a helical ribbon}
\label{sec:geometry}

A ribbon is defined as an elastic strip of length $L$, width $w \ll L$, and thickness $H \ll w$ (Chen et al. 2011)
.  The cross-section of the ribbon is rectangular and the principle geometric axes of the ribbon are oriented along the ribbon length ($\dx$), width ($\dy$), and thickness ($\dz$) directions.  These directions form an orthonormal triad of vectors $\{\dx,\dy,\dz\}$ that rotate with the ribbon as it coils in a three-dimensional Cartesian frame $\{\Ex,\Ey,\Ez\}$. This representation originates from the directed rod theory in Green \& Laws (2006) and OReilly (1998)

In planar bending (Stoney 1909)
, curvature is restricted to the $\dx$ (length-wise) direction. In the general case of helical deformation, illustrated in Fig.~\ref{fig:helix}, curvature will occur along principle directions
\begin{equation}
\rx = \dx \cos\phi - \dy \sin\phi \andsp \ry = \dx \sin\phi + \dy \cos\phi \,,
\end{equation}
where $\phi$ represents the orientation of the principle curvatures with respect to the principle geometric axes and $\kappa_1$ and $\kappa_2$ denote the principle curvatures along the directors $\rx$ and $\ry$, respectively.  Points along the centerline of the ribbon follow a space curve
\begin{equation}
\textbf{P} = \textbf{P}_0(s) = X_0(s) \Ex + Y_0(s) \Ey +  Z_0(s) \Ez\,,
\end{equation}
where $s \in \{0,L\}$ is the convecting arclength.  In the following, we derive a relationship between the ribbon curvatures and orientation $\{\kappa_1,\kappa_2, \phi\}$ and the pitch, radius, axis, and chirality of the helical space curve $\bP = \bP_0(s)$.  In doing so, it is convenient to define the unit normal $\bN = \dz \equiv \dx\times\dy = \rx\times\ry$.  By definition,
\begin{equation} \label{curvatures}
\Dfrac{\rx}{s} = - \bN \kappa_1 \cos \phi
\andsp \Dfrac{\ry}{s} = - \bN \kappa_2 \sin \phi \,.
\end{equation}
Lastly, the identities $\mathrm{d}{\bP}/\mathrm{d}s = \dx$ and $\bN = \rx\times\ry$ together imply
\begin{equation} \label{identities}
\Dfrac{\bP}{s} =  \sin{\phi} \rx  + \cos{\phi}  \ry \andsp \Dfrac{\bN}{s} =  \kappa_1 \cos{\phi} \rx + \kappa_2 \sin{\phi} \ry \,.
\end{equation}

\subsection{Derivation of the helix centerline coordinates} \label{sec:P}

Next, we derive a differential equation for ${\bN}(s)$ and solve it explicitly.  To this end, from Eqs.~(\ref{curvatures}) and (\ref{identities}) it directly follows that
\begin{equation}
\Dfrac{^2 \bN}{s^2} = - \left(\kappa_1^2 \cos^2 \phi + \kappa_2^2 \sin^2 \phi\right) \bN \,.
\label{Diff^2_N}
\end{equation}
This second-order differential equation is subject to the initial conditions $\bN(0) = \Ez$ and
\begin{equation}
\Dfrac{\bN(0)}{s} =  \kappa_1 \cos\phi \rx(0) + \kappa_2 \sin\phi \ry(0) \,,
\end{equation}
where
\begin{equation} \label{curvature_IC}
\rx(0) = \cos\phi\Ex - \sin\phi\Ey \andsp \ry(0) = \sin\phi\Ex + \cos\phi\Ey\,.
\end{equation}
Solving for $\bN(s)$ yields
\begin{equation}
\bN = \left(\frac{\beta}{\alpha} \sin \alpha s\right) \Ex - \left(\frac{\tau}{\alpha} \sin \alpha s\right) \Ey  + \left(\cos \alpha s\right) \Ez \,,
\label{N_Sol}
\end{equation}
where
\begin{equation}
\alpha = \sqrt{\kappa_1^2 \cos^2 \phi + \kappa_2^2 \sin^2 \phi},
\label{eq:alpha}
\end{equation}
\begin{equation}
\beta = \kappa_1 \cos^2 \phi + \kappa_2 \sin^2 \phi,
\label{eq:beta}
\end{equation}
and
\begin{equation}
\tau = (\kappa_1 - \kappa_2)\sin \phi \cos \phi \,.
\end{equation}
In passing, we note that $\alpha^2= \beta^2 + \tau^2$.
Next, integrating Eqs.~(\ref{curvatures}) for the initial conditions expressed in Eq.~(\ref{curvature_IC}) yields
\begin{eqnarray}
\rx &=& \left\{\cos \phi + \frac{\beta}{\alpha^2} \kappa_1 (\cos \alpha s - 1) \cos \phi\right\} \Ex \nonumber\\
      & &  + \left\{-\sin \phi - \frac{\tau}{\alpha^2} \kappa_1 (\cos \alpha s - 1) \cos \phi\right\} \Ey \nonumber\\ & & - \left\{\frac{\kappa_1}{\alpha} \sin \alpha s \cos \phi \right\}) \Ez \\
\ry &=& \left\{\sin \phi + \frac{\beta}{\alpha^2} \kappa_2  (\cos \alpha s - 1) \sin \phi\right\} \Ex \nonumber\\
& & + \left\{\cos \phi - \frac{\tau}{\alpha^2} \kappa_2 (\cos \alpha s - 1) \sin \phi\right\} \Ey \nonumber\\ & & - \left\{\frac{\kappa_2}{\alpha}  \sin \alpha s \sin \phi\right\} \Ez\,.
\end{eqnarray}
Lastly, integrating the expression for $\mathrm{d}{\bP}/\mathrm{d}s$ in Eq.~(\ref{identities}) implies
\begin{eqnarray}
X_0(s) &=& s - \frac{\beta^2}{\alpha^3} (\alpha s - \sin \alpha s) \nonumber\\
Y_0(s) &=&  \frac{\beta \tau}{\alpha^3} (\alpha s - \sin \alpha s) \nonumber\\
Z_0(s) &=&  \frac{\beta}{\alpha^2} (\cos \alpha s - 1) \,.
\label{eq:kinematics}
\end{eqnarray}
As a consistency check, we see that  $\kappa_1=\kappa_2=0$, $X_0(s)=s$ while $Y_0(s)=Z_0(s)=0$ describes a straight ribbon oriented along the $\Ex$ direction.

\subsection{Geometric properties of the helical space curve $\bP$}

The space curve $\bP$ derived in Sec.~\ref{sec:P} represents a helix with curvature $\beta$, torsion $\tau$, axis ${\bf M}$, helix angle $\Phi$ (the angle between a tangent vector of the helix and the plane normal to the helix axis), radius $R$, and pitch $D$.   The corresponding Frenet-Serret frame $\{ \textbf{T}, \textbf{N}, \textbf{B} \}$ (Tu and Ou-Yang 2008)
satisfies the following differential equations:
\begin{equation}
\Dfrac{\textbf{T}}{s} = -\beta \textbf{N}(s) \qquad
\Dfrac{\textbf{N}}{s} = \beta \textbf{T}(s) + \tau \textbf{B}(s) \qquad
\Dfrac{\textbf{B}}{s} = -\tau \textbf{N}(s) \,.
\label{Eqn_FrenetFrame}
\end{equation}
Here, $\textbf{T} = \dx$ is the tangent vector,
\begin{equation}
\textbf{T}(s) =  \left\{1 - \frac{\beta^2}{\alpha^2}(1 - \cos \alpha s)\right\} \Ex + \frac{\beta \tau}{\alpha^2}(1 - \cos \alpha s)  \Ey  - \left\{\frac{\beta}{\alpha} \sin \alpha s \right\} \Ez ,
\label{T_Sol}
\end{equation}
and $\textbf{B} = {\bf T}\times\bN$ (Kamien 2002)
 is a bi-normal vector:
\begin{equation}
\textbf{B} = \frac{\beta \tau}{\alpha^2}(1 - \cos \alpha s) \Ex + \left\{\frac{\beta^2}{\alpha^2}(1 - \cos \alpha s) + \cos \alpha s\right\} \Ey  + \left\{\frac{\tau}{\alpha} \sin \alpha s\right\} \Ez ,
\label{B_Sol}
\end{equation}

Now, the axis ${\bf M}$ is parallel to the vector formed by one complete repeating unit: $\textbf{P}(2 \pi/\alpha) - \textbf{P}(0) =2 \frac{\pi \tau}{\alpha^2} \left\{\frac{\tau}{\alpha} \Ex + \frac{\beta}{\alpha} \Ey\right\}$. Normalization yields ${\bf M} =  \frac{\tau}{\alpha} \Ex + \frac{\beta}{\alpha} \Ey$, implying that, for a helix, $\textbf{T} \cdot \textbf{M} = \tau/\alpha$ is a fixed value, which corresponds to the helix angle
\begin{equation} \label{angle}
\Phi = \arcsin \left(\frac{\tau}{\alpha}\right) = \arctan \left\{\frac{(\kappa_1 - \kappa_2) \sin \phi \cos \phi}{\kappa_1 \cos^2{\phi} + \kappa_2 \sin^2{\phi}}\right\}\,.
\end{equation}
Helical chirality is determined by the sign of the helix angle (or equivalently the torsion of the ribbon centerline), i.e., $ \mathop{\mathrm{sgn}}(\Phi)$.  Right handed helices correspond to  $\Phi > 0$ whereas left-handed helices correspond to $\Phi < 0$.   Equation~(\ref{angle}) implies that the chirality is determined by the signs of both $(\kappa_1 - \kappa_2)$ and $\sin \phi \cos \phi$.  As expected, when interchanging the values of $\kappa_1$ and $\kappa_2$, the helix shape transforms to its mirror image about the plane $y = z$ with reversed handedness.  Lastly, the helix has a radius
\begin{equation}
R = \frac{\beta}{\alpha^2} \frac{\kappa_1 \cos^2{\phi} + \kappa_2 \sin^2{\phi}}{\kappa_1^2 \cos^2{\phi} + \kappa_2^2 \sin^2{\phi}}
\label{equation:R}
\end{equation}
and pitch (\footnote{An alternative method for deriving $R$ and $D$ is to examine Eq.~(\ref{Eqn_FrenetFrame}) directly. Since the Frenet-Serret triad associated with the centerline satisfies Eq.~(\ref{Eqn_FrenetFrame}), with constant curvature $\beta = \kappa_1 \cos^2 \phi + \kappa_2 \sin^2 \phi$ and torsion $\tau = (\kappa_1 - \kappa_2)\sin \phi \cos \phi$, the centerline is indeed a circular helix with constant radius $R = \beta/(\beta^2 + \tau^2) = \beta/\alpha^2$ and pitch $D = 2 \pi \tau/\alpha^2$.})
\begin{equation}
D = ||\textbf{P}(2 \pi/\alpha) - \textbf{P}(0)|| =  \frac{2 \pi \tau}{\alpha^2}\,.
\end{equation}

\begin{figure}[!h]
\graphicspath{{figs/}}
\includegraphics[width=5in]{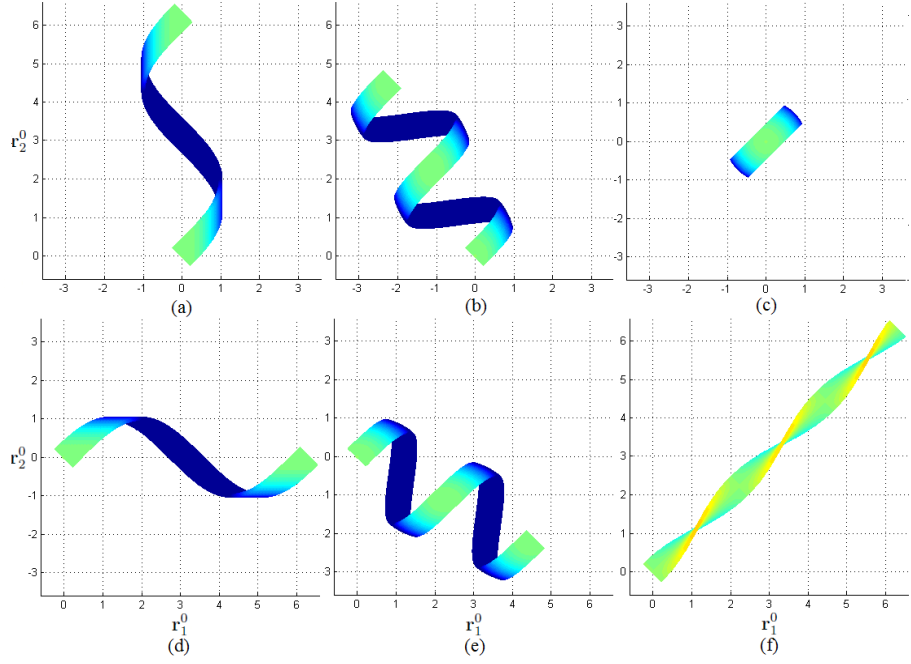}
\caption{Spectrum of representative ribbon morphologies as relative values of $\kappa_1$ and $\kappa_2$ are tuned ($\phi = \pi/4$ in all these cases). As described in Section \ref{sec:geometry}, the ribbon originally lies along $\Ex$ direction in global coordinate system (which forms an angle $\phi$ from ${\bf r}^0_1$), in the plane spanned by ${\bf r}^0_1$ and ${\bf r}^0_2$ where ${\bf r}^0_1 = \cos \phi \Ex -\sin \phi \Ey$ and ${\bf r}^0_2 = \sin\phi \Ex + \cos\phi \Ey$. The values of principal curvatures $\kappa_1$ and $\kappa_2$ are as follows: (a) $\kappa_1 = 1, \kappa_2 = 0$;  (b) $\kappa_1 = 1, \kappa_2 = 1/2$ ; (c) $\kappa_1 = \kappa_2 = 1$; (d) $\kappa_1 = 0, \kappa_2 = 1$; (e) $\kappa_1 = 1/2, \kappa_2 = 1$ ; (f) $\kappa_1 = -1, \kappa_2 = 1$ .}
\label{Kinematics_study} 
\end{figure}

\subsection{Helical space curve:  Special cases}

We now consider a few special helical shapes before describing the more general case. The ribbon forms a ring when either $\kappa_1 = \kappa_2$, $\phi = 0$, or $\phi = \pi/2$, which leads to  helix angle of zero according to Eq.~(\ref{angle}). For example, when $\kappa_1 = \kappa_2$ [as shown in Fig.~\ref{Kinematics_study}$(c)$], the helix angle is $\Phi = 0$ and the helix reduces to a ring. A ring can also be obtained when $\phi = 0$ or $\pi/2$. Moreover, when the Gauss curvature $K = \kappa_1 \kappa_2$ is zero, a helical cylindrical ribbon is formed. For example, setting $\kappa_2 = 0$ [cf.~Fig.~\ref{Kinematics_study}$(a)$] yields a right-handed helical ribbon that wraps around a cylinder of radius $1/\kappa_1$, while setting $\kappa_1 = 0$ [cf.~Fig.~\ref{Kinematics_study}$(d)$] leads to a left-handed helical ribbon wrapping around a cylinder of radius $1/\kappa_2$. Another configuration of particular interest is the purely twisted ribbon, which forms when $\kappa_1 \cos^2{\phi} + \kappa_2 \sin^2{\phi} = 0$ implying that the helix angle is $\Phi = \pi/2$ and radius $R = 0$. A specific example is shown in Fig.~\ref{Kinematics_study}$(f)$ for the case $\kappa_2 = -\kappa_1$ (i.e., the mean curvature $H=(\kappa_1 + \kappa_2)/2=0$) and $\phi =\pi/4$.  Here, the ribbon undergoes a pure twist deformation (i.e., the centerline remains straight).  In other words, by tuning the values of $\kappa_1$, $\kappa_2$ (or equivalently the mean and Gauss curvatures) and  $\phi$ (in the above cases, $\phi$ is fixed to be $\pi/4$, but in general, $\phi$ is another tunable geometric parameter), enables one to generate helical shapes of any kind.  More specifically, transitions between rings, cylindrical helical ribbons, and purely twisted ribbons can be achieved by tuning the set of three independent geometric parameters, namely the two principal curvatures $(\kappa_1,\kappa_2)$ and the misalignment $\phi$ between principal curvature and geometric axes.  For given $\kappa_1$, $\kappa_2$, and $\phi$, the ribbon morphologies were generated by the procedure outlined in the Appendix.

\section{Continuum elasticity theory and ribbon morphology}

Having established a quantitative link between prescribed ribbon principal curvatures and ribbon morphology in Sec.~\ref{sec:geometry}, we now focus on the role of surface stress in controlling the principal curvatures (and hence morphology).   To this end, we determine the equilibrium shape of the helical ribbon by minimizing its elastic strain energy,
\begin{equation}
\Pi = {\bf f}^-:\strain |_{z=-H/2} + {\bf f}^+:\strain |_{z=H/2}  + \int_{-H/2}^{H/2}\frac{1}{2}\strain:{\bf C}:\strain\,dz\,.
\end{equation}
Here, $\strain$ is the strain tensor for linear elastic deformation, ${\bf C}$ is the fourth-order elastic stiffness tensor, and $\{{\bf f}^+,{\bf f}^-\}$ are the surface stresses.  The coordinate $z\in\{-H/2,H/2\}$ is defined through the thickness of the ribbon and ${\bf f}^+$ and ${\bf f}^-$ corresponding to the surfaces at $z = H/2$ and $z = -H/2$, respectively.

\subsection{Kinematics and stationarity principle}

As in classical plate theory, the strain tensor $\strain$ is obtained by superimposing strains induced by elastic bending ($\strain_b$) and an in-plane uniform ``membrane'' strain ($\strain_m$).  We also include strain along $\dz$ in order to allow for plane stress compatibility ($\strain_z$).  In general, the ribbon may also be subject to internal residual strains that do not correspond to surface stress ($\strain_0$).  Combining these contributions leads to a strain tensor of the form
\begin{equation}
\strain = \strain_b + \strain_m + \strain_z + \strain_0 \,,
\end{equation}
where
\begin{eqnarray}
\strain_b &=& z \kappa_1 \rx \otimes \rx + z \kappa_2 \ry \otimes \ry \nonumber\\
\strain_m &=& \varepsilon_{xx}\dx\otimes\dx + \varepsilon_{xy}(\dx\otimes\dy + \dy\otimes\dx) + \varepsilon_{yy}\dy\otimes\dy \nonumber\\
\strain_z &=& (\varepsilon_{zz} + qz)\dz\otimes\dz \nonumber\\
\strain_0 &=& \gamma_{xx}^0\dx\otimes\dx + \gamma_{xy}^0(\dx\otimes\dy + \dy\otimes\dx)  + \gamma_{yy}^0\dy\otimes\dy + \gamma_{zz}^0\dz\otimes\dz \,.
\end{eqnarray}
Here, $q$ denotes the gradient of the strain component along $\dz$ required for plane stress compatibility.

The strain tensor may also be expressed as $\strain = \gamma_{ij}{\bf d}_i\otimes{\bf d}_j$ ($i,j \in \{x,y,z\}$) with components
\begin{eqnarray}
\gamma_{xx} &=& \epsilon_{xx} + z (\kappa_1 \cos^2 \phi + \kappa_2 \sin^2 \phi)+\gamma_{xx}^0(z) \nonumber\\
\gamma_{xy} &=& \gamma_{yx} = \epsilon_{xy} + z (\kappa_2 - \kappa_1) \sin \phi \cos \phi+\gamma_{xy}^0(z) \nonumber\\
\gamma_{yy} &=& \epsilon_{yy} + z (\kappa_1 \sin^2 \phi + \kappa_2 \cos^2 \phi)+\gamma_{yy}^0(z) \nonumber\\
\gamma_{zz} &=& \epsilon_{zz} + qz+\gamma_{zz}^0(z) \,.
\end{eqnarray}

Let $\phi^+$ and $\phi^-$  denote the orientation of the principle axes of the surface stress on the top ($z = H/2$) and bottom ($z = -H/2$) surfaces, respectively.  Unlike $\phi$, which is an unknown geometric parameter, both $\phi^+$ and $\phi^-$ represent fixed values, prescribed by nature or in manufacturing. For example, they could coincide with the principle axes of the surface stress tensor, ${\bf f}$.  On the top and bottom ($z = {\pm}H/2$) surfaces, ${\bf f}$ adopts the following forms: ${\bf f}^{\pm}= f_1^{\pm} {\bf e}_1^{\pm}\otimes{\bf e}_1^{\pm} + f_2^{\pm} {\bf e}_2^{\pm}\otimes{\bf e}_2^{\pm}$, where ${\bf e}_1^{\pm} = \cos\phi^{\pm}\dx - \sin\phi^{\pm}\dy$ and ${\bf e}_2^{\pm} = \sin\phi^{\pm}\dx + \cos\phi^{\pm}\dy$.

At equilibrium, the variational principle dictates that $\Pi$ must be stationary with respect to the variations of all the unknown parameters $\kappa_1$, $\kappa_2$, $q$, $\epsilon_{xx}$, $\epsilon_{yy}$, $\epsilon_{xy}$, $\epsilon_{zz}$ and $\phi$.  In other words, these values correspond to the solutions to the following set of linearly independent equations: $\partial\Pi/\partial\chi = 0$, where $\chi$ represents any one of the eight unknowns.  It is worth noting that although here we choose to write the strain tensor in the $\{\dx,\dy,\dz\}$ coordinate system, it could also be expressed in other orthonormal coordinate systems, such as $\{{\bf r}_1, {\bf r}_2, \dz\}$ (which we will adopt in the example that follows). The total potential energy density, however, remains the same regardless of which orthonormal coordinate system one chooses.

\subsection{Analytical solutions}

In general,  the system of eight equations $\partial\Pi/\partial\chi = 0$ for the eight unknowns $\{\chi\}$ can only be obtained numerically.  However, closed-form algebraic solutions are possible when the ribbon is treated as homogenous, elastically isotropic, and subject to surface stress on only one surface. Without loss of generality, we consider surface stress acting on the bottom surface ($z = -H/2$) only:  ${\bf f}^- = f_1^- {\bf e}^-_1 \otimes {\bf e}^-_1 + f_2^- {\bf e}^-_2 \otimes {\bf e}^-_2$, and the angle from $\dx$ to ${\bf e}^-_1$ (clockwise) is $\phi^-$. As illustrated in Fig.~\ref{fig:helix}, the strain tensor is diagonalized in the orthonormal coordinate system $\{{\bf r}_1, {\bf r}_2, \dz\}$, so that the only nonzero strain components are $\gamma_{11} = \epsilon_{11} + z \kappa_{1}$, $\gamma_{22} = \epsilon_{22} + z \kappa_{2}$ and $\gamma_{33} = \epsilon_{33} + q z$. Thus, the total potential energy per unit area is
\begin{eqnarray}
\Pi &=& \frac{EH^3}{24} \left\{\frac{\nu (\kappa_{1} + \kappa_{2} + q)^2}{(1-2\nu)(1+\nu)} + \frac{ (\kappa^2_{1} + \kappa^2_{2} + k^2_{3}) }{1+\nu}\right\} \nonumber \\
&+& \frac{EH}{2}\left\{\frac{\nu (\epsilon_{11} + \epsilon_{22} + \epsilon_{33})^2 }{(1-2\nu)(1+\nu)} + \frac{(\epsilon^2_{11} + \epsilon^2_{22} + \epsilon^2_{33})}{1+\nu}\right\} \\
     &+&  \left\{f_1^- \cos^2 (\phi - \phi^-) + f_2^- \sin^2 (\phi - \phi^-) \right\} \left(\epsilon_{11} - \frac{\kappa_1 H}{2}\right) \nonumber\\
     &+&  \left\{f_1^- \sin^2 (\phi - \phi^-) + f_2^- \cos^2 (\phi - \phi^-) \right\}  \left(\epsilon_{22} - \frac{\kappa_2 H}{2}\right) \, .\nonumber
\end{eqnarray}
where $E$ and $\nu$ denote the Young's modulus and Poisson's ratio of the ribbon, respectively.

Applying the variational principles implies that the principle directions of the curvature and surface stress coincide, i.e., $\phi = \phi^-$, and that
\begin{eqnarray}
\kappa_1 &=& \frac{6(f_1^- - \nu f_2^-)}{E H^2} \qquad
\kappa_2 = \frac{6(f_2^- - \nu f_1^-)}{E H^2} \nonumber\\
q \,\, &=& -\frac{6\nu (f_1^- + f_2^-)}{E H^2} \,\,\,\,\,
\epsilon_{11} = -\frac{f_1^- - \nu f_2^-}{E H} \nonumber\\
\epsilon_{22} &=& -\frac{f_2^- - \nu f_1^-}{E H} \qquad
\epsilon_{33} = \frac {\nu(f_1^- + f_2^-)}{E H}\, .
\end{eqnarray}
It is interesting to note that in the absence of Poisson coupling, the first two equations are reduced to the classical Stoney formula (Stoney 1909)
in the two principle curvature directions:  $\kappa_1 = 6f_1^-/E H^2$ and $\kappa_2 = 6f_2^-/E H^2$.

For the more general case, where surface stresses occur on both surfaces, we can decouple the stretching and bending modes. First, we define the stretching component of the surface stress ${\bf f}^{s} = {\bf f^+} + {\bf f^-}$ and the bending component ${\bf f}^* = {\bf f^-} - {\bf f^+}$, where ${\bf f}^*$ represents the effective surface stress acting on the bottom surface ($z = -H/2$).  Next, ${\bf f}^{s}$ and ${\bf f}^*$ can be diagonalized with respect to the principle axes $(\dux, \duy)$ and $(\drx,\dry)$: ${\bf f^{s}} = f^{s}_1 \dux \otimes \dux + f^{s}_2 \duy \otimes \duy$ and ${\bf f}^* = f_1^* \drx \otimes \drx + f^*_2 \dry \otimes \dry$. Thus, by de-coupling the stretching and bending modes, and applying variational principles, the total strain tensor $\strain = \strain_s + \strain_b$ can be obtained.$\strain_s$ is the strain component due to stretching
\begin{equation}
\strain_s = - \frac{f^{s}_1 - \nu f^{s}_2}{E H} \dux \otimes \dux - \frac{f^{s}_2 - \nu f^{s}_1}{E H} \duy \otimes \duy  + \frac{\nu(f^{s}_1 + f^{s}_2)}{E H} \dz \otimes \dz
\label{stretching_tensor}
\end{equation}
and $\strain_b$ is the strain component caused by an effective surface stress on the bottom surface
\begin{equation}
\strain_b =  \kappa_1  z \drx \otimes \drx + \kappa_2 z \dry \otimes \dry +  q z\dz \otimes \dz \,.
\label{bending_tensor}
\end{equation}
Thereby, the problem is reduced to an equivalent system with surface stress only on the bottom surface, with $\kappa_1 = 6(f_1^* - \nu f_2^*)/E H^2$, $\kappa_2 = 6(f_2^* - \nu f_1^*)/E H^2$, and $q = -6\nu (f_1^* + f_2^*)/E H^2$.

\section{Experimental validation and discussion}

\begin{figure}[t]
\begin{center}
\includegraphics[width=3.5in]{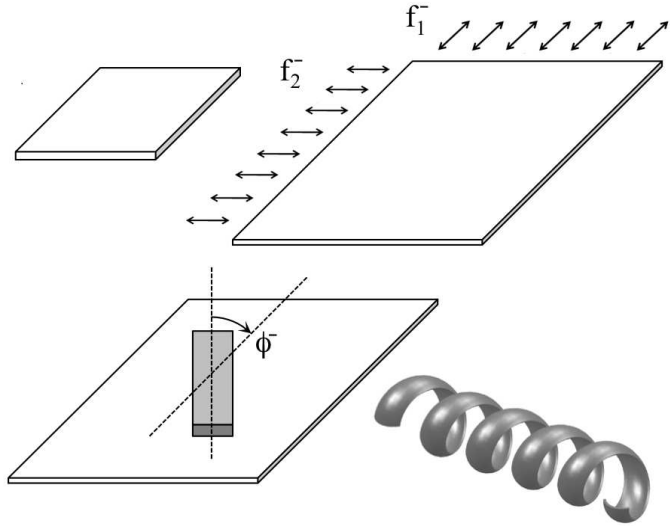}
\end{center}
\caption{Fabrication of a helical ribbon.  A strip of elastic, pressure sensitive adhesive is bonded to a pre-stretched sheet of latex rubber, with a misorientation angle $\phi^{-}$ between the principal curvature and geometric axes.  }
\label{fig:experiment}
\end{figure}

\begin{figure}[!h]
\begin{center}
\includegraphics[width=3in]{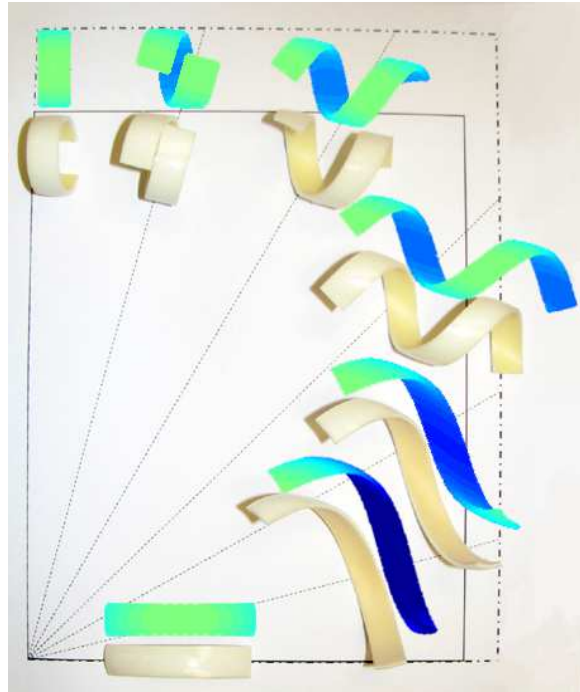}
\end{center}
\caption{A square piece of latex rubber (solid lines) is pre-stretched twice as much in the vertical direction than in the horizontal direction (to the dash-dot edges). An elastic adhesive sheet with no pre-strains is then bonded to the strained latex sheet and subsequently cut into strips with long axes varying between $0^{o}$ and $90^{o}$ (at an interval of $15^{o}$).  The relaxed ribbons are shown at the appropriate angles along with the corresponding theoretical predictions.
}
\label{fig:results}
\end{figure}

\begin{figure}[!h]
\begin{center}
\includegraphics[width=3in]{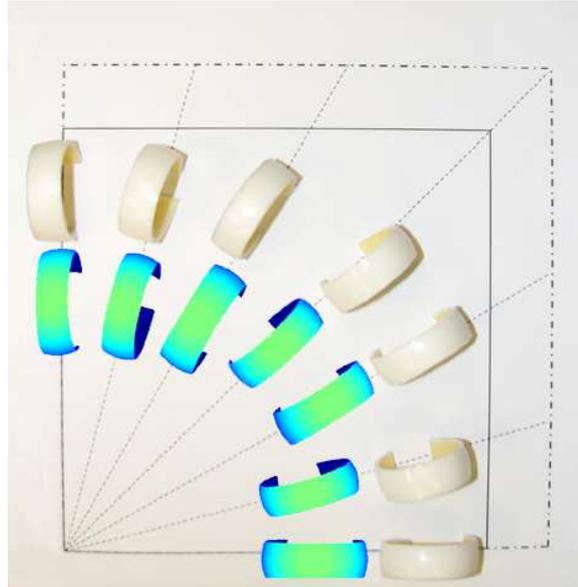}
\end{center}
\caption{A square piece of latex rubber (solid lines) is stretched equally in the vertical and horizontal direction (dash-dot edges). An unstrained elastic adhesive sheet is then bonded to the strained latex sheet and subsequently cut into ribbons with long axis varying between $0^{o}$ and $90^{o}$ ($15^{o}$ intervals).  The released samples are shown at the appropriate angles along with the corresponding theoretical predictions.  All of the ribbons adopt ring shapes with identical radii regardless of the orientation.
}
\label{fig:results2}
\end{figure}

\begin{figure}[!h]
\begin{center}
\centerline{\includegraphics[height=3in]{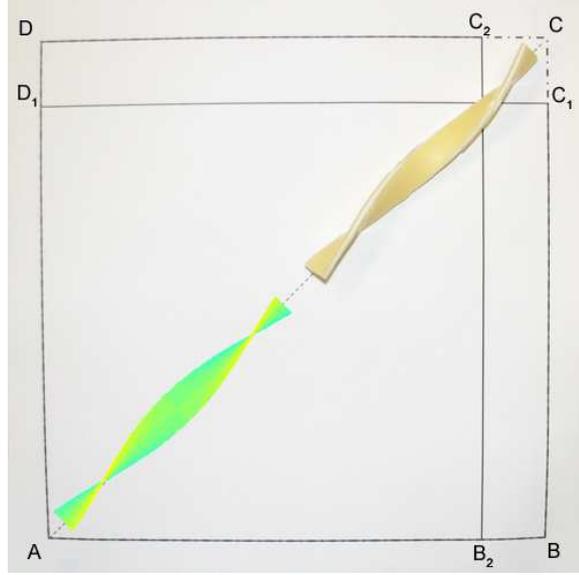}}
\end{center}
\caption{A rectangular bottom sheet ($A B C_1 D_1$) and a top sheet ($A B_2 C_2 D$) are pre-stretched in the vertical and horizontal direction, respectively, so that they both become a square sheet ($A B C D$), before bonding with an unstrained elastic strip with $\phi = 45^{o}$. As predicted, the strip deforms into a purely twisted ribbon.}
\label{fig:results3}
\end{figure}

We validated our theoretical predictions with a series of simple, table-top experiments in which a thin sheet of latex rubber 
was pre-stretched and bonded to an elastic strip of thicker, pressure-sensitive adhesive (Fig. \ref{fig:experiment}), similar to the physical models employed by the current and other researchers (Chen et al. 2011; Armon et al. 2011; Chen et al. 2012; Huang et al., 2012; Gerbode et al. 2012)
.  The latex sheets were purchased from {\it Small Parts Inc.} and the elastic strips were Acrylic, Scotch Wall-Mounting Tape, produced by {\it 3M}. Their Young's moduli were $1.4$MPa and $10.3$MPa respectively, measured by an Instron$^{TM}$ 5848 Micro-Tester, and the thicknesses were $0.048$cm and $0.1$cm respectively. The Poisson's ratio of Acrylic is $0.37$ (Powers and Caddell 1972)
and that of the latex sheets is $0.49$ (Kaazempur-Mofrad et al. 2003)
. Helical ribbons with different helix angle, radius, axis, chirality, and Gauss curvature were obtained by altering the magnitudes of the two principle pre-stretches and their angle with respect to the centerline orientation of the bonded strip.  These results were then compared with the theoretical predictions, as described in the captions to Figs.~\ref{fig:results}-\ref{fig:results3}. In Figs.~\ref{fig:results}, the predicted principal radii of curvatures are $19.6cm$ and $123cm$ respectively, consistent with the experimentally measured values $17.6cm$ and $112cm$. In Fig. \ref{fig:results2}, the predicted principal radius of curvature is $27.0cm$, in agreement with the experimental value $23.0cm$. While for the purely twisted ribbon as shown in Fig. \ref{fig:results3}, the theoretical calculation and the experimentally measured value of the principal radii of curvature are $\pm 12.4cm$ and $\pm 13.6cm$ respectively, again showing reasonable agreement.

When the pre-stretches are anisotropic, helical ribbons can exhibit a broad range of helix angles, radii, and  centerline orientations. As shown in Fig.~\ref{fig:results}, the theoretical predictions for helix angle and ribbon shapes are in good agreement with the experimental observations with no adjustable parameters.  We note that experimental validation of the theory has also been demonstrated for the more general case in which both sides of the elastic strip are bonded to differently oriented pre-stretched layers of latex rubber (Chen et al. 2011)

It is important to emphasize that helicity in these systems arises from mechanical anisotropy.  Such anisotropy may originate from elastic anisotropy due to molecular tilt or chiral interactions (Selinger et al. 2004)
, or mechanical anisotropy in surface/external stresses, residual strains and elastic properties (Zhang et al. 2005; Wang et al. 2008; Dai and Shen 2009; Pokroy et al. 2009)
. In molecular dynamics simulations, anisotropy can also arise from the asymmetry in the direction along which periodic boundary condition is applied (Lee et al. 2011)
. The presence of at least one of these  sources of anisotropy is required in order to break the geometric symmetry and define the chirality (handedness) of the helical ribbon. For single-layered ribbons, helicity can be introduced by the anisotropy of the in-plane principle components of either the elastic stiffness or surface stress tensors, or a combination of both.

Although our current model only accounts for the effects of surface stress anisotropy, the same principle can be applied to treat other types of anisotropy that lead to helicity, such as helicity in ribbon laminates composed of two or multiple layers due to lattice mismatch or pre-strained individual layers (Zhang et al. 2005)
. Alternatively, if the pre-stretches are isotropic (biaxial tension), ring shapes will result, regardless of the different centerline orientations, as often observed in the literature (Mei et al. 2009; Cho et al. 2010)
. Figure~\ref{fig:results2} shows good agreement between the experimental results and theoretical predictions with such equal-biaxial pre-stretches on the bottom elastic sheet.

Purely twisted ribbons represent yet another class of interesting morphologies (Oda et al. 1999; Selinger et al. 2001; Liu et al. 2009)
. Such a deformation is outside the scope of developable ribbon deformations (Giomi and Mahadevan 2010)
, since the deformation of a purely twisted ribbon is non-isometric (i.e., local distances between points are not always preserved). Since our theory takes into account all possible deformation modes (i.e., stretching, bending and twisting) of the ribbon, it has the capability of capturing pure twist. As predicted by the current theory, when $\kappa_1 \cos^2{\phi} + \kappa_2 \sin^2{\phi} = 0$, $\Phi = \pi/2$,  a purely twisted ribbon will form. This was also verified in our experiments, as shown in Fig.~\ref{fig:results3}. The rectangular bottom sheet ($A B C_1 D_1$) and the top sheet ($A B_2 C_2 D$) are pre-stretched in the vertical and horizontal direction respectively so that they both become a square sheet ($A B C D$), before bonded with an unstrained elastic strip with $\phi = 45^{o}$. According to our theory, this yields $\kappa_2 = -\kappa_1$ (the same as shown in Fig.~\ref{Kinematics_study}) resulting in a purely twisted ribbon; the predicted morphology is in excellent agreement with the corresponding experimental result.

\begin{figure}[!h]
\graphicspath{{figs/}}
\centerline{\includegraphics[height=4in]{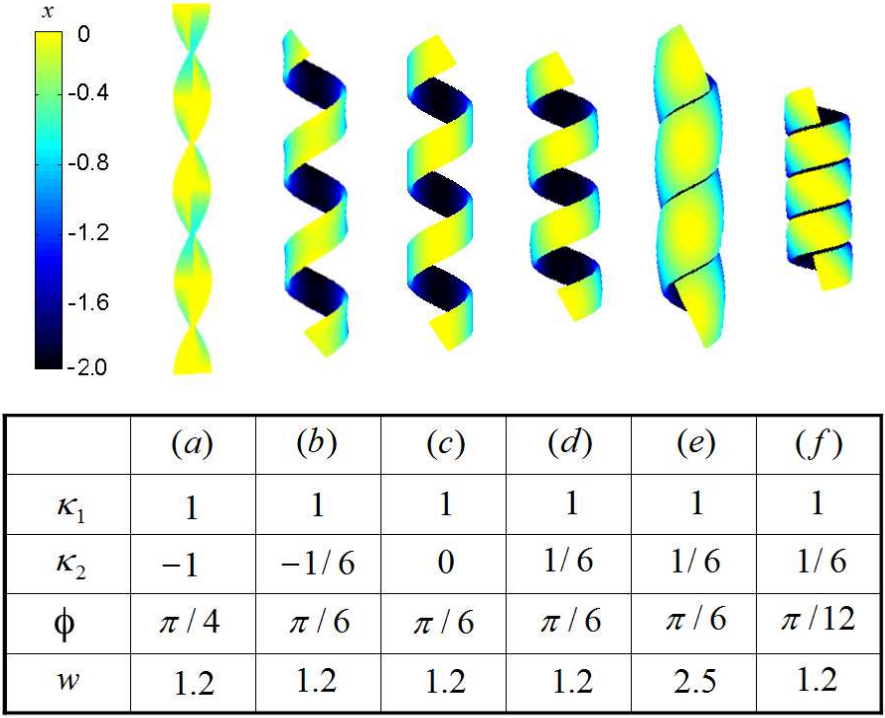}}
\caption{Structural transition from twisted to helical ribbons and tubules with varying principal curvatures $(\kappa_1,\kappa_2)$, misorientation $\phi$, and width $w$.}
\label{StructuralTransition}
\end{figure}

Structural transitions among cylindrical helical ribbons, general helical ribbons, purely twisted ribbons, and cylinders/tubules are all captured within the current theoretical framework, as illustrated in Figs.~\ref{Kinematics_study} and \ref{StructuralTransition}. Similar transitions have been observed in bio-chemical systems such as cholesterol (Chung et al. 1993)
, surfactants (Oda et al. 1999)
and peptides (Bellesia et al. 2008; Pashuck and Stupp 2010)
, as well as in inorganic materials such as the transition metal dichalcogenide $WS_2$ (Iglic et al. 2005) 
. Understanding these transitions is key to programmable fabrication of nano- and bio-devices, and possibly, a better understanding of certain diseases where morphological changes of helical ribbons and tubules are key (Chung et al. 1993)
. More specifically, our kinematic model shows that when the Gauss curvature $K$ is negative,  the ribbon takes on a local saddle shape (the first two panels in Fig.~\ref{StructuralTransition}), and in particular, if $\kappa_1 \cos^2{\phi} + \kappa_2 \sin^2{\phi} = 0$ , the ribbon becomes a purely twisted strip (the first panel in Fig.~\ref{StructuralTransition}) as investigated by (Selinger et al. 2001)
, (Ghafouri and Bruinsma 2005)
, and (Igli$\acute{c}$ et al. 2005)

When the Gauss curvature vanishes, a helical ribbon with cylindrical curvature (like a ribbon bounding exactly about a cylinder so that the edges are straight) adopts the third configuration in Fig.~\ref{StructuralTransition}. However, for non-zero mean curvature and non-zero Gauss curvature, the general configuration is still helical, as shown in the second and fourth configurations of Fig.~\ref{StructuralTransition}.  For these ribbons, the edges are not exactly straight and could be either concave or convex depending on the sign of the Gauss curvature. This represents a class of general helical configurations that have been largely overlooked in the literature but may contribute to the understanding of the gradual conformational changes from twisted to helical ribbons that occur in a variety of physical systems (Fuhrhop and Helfrich 1993, Oda et al. 1999, Selinger et al. 2004) 
. On the other hand, the last two configurations in Fig.~\ref{StructuralTransition} show the formation of ``tubules'' when the edges of the ribbon touch;  such configurations can be obtained by either increasing ribbon width $w$ at fixed misorientation $\phi$ or by varying $\phi$ at fixed $w$. The as-generated tubules are similar to those formed in chiral liquid crystals due to molecular tilt modulation as illustrated in Fig.~6 of (Selinger et al. 2001)
, although the origins of the mechanical anisotropies are presumably different.

Historically, there have been limited theoretical attempts to explain the change between purely twisted ribbons with saddle curvature ($K < 0$) and helical cylindrical geometries ($K = 0$) (Bellesia et al. 2008; Ghafouri and Bruinsma 2005; Igli$\acute{c}$ et al. 2005; Selinger et al. 2001; Selinger et al. 2004)
. The existing models often attribute the morphological changes to the asymmetry in the chemical nature and the affinity of the two molecular surfaces to the solvent, or the force differences on two surfaces that lead to preferential bending or twisting. These asymmetric conditions, when applied in molecular dynamics or Monte Carlo simulations, yield configurations consistent with experiments (Bellesia et al. 2008; Selinger et al. 2004)
. Nevertheless, these approaches are computationally expensive, and finding an energy minimum often requires a starting configuration near the minimum energy state for fast convergence. In contrast, the current theory, by conveniently parameterizing those asymmetries through surface stress or elasticity, provides a continuum approach towards quantitatively understanding the associated structural transitions induced by tuning the relevant geometric parameters. Of course, proper coarse-graining models need to be developed to establish the link between the asymmetry/anisotropy in molecular interactions and the effective, anisotropic surface stresses that result in helicity.

\section{Conclusions}

A mathematical analysis has been developed to predict the formation of families of helices, rings, and twisting ribbons based upon three independent geometric parameters ($\kappa_1$, $\kappa_2$, $\phi$).  Continuously varying these parameters allows for a smooth transition between the different shape classes.  This model employs continuum elasticity, differential geometry, and stationarity principles to predict how a naturally flat ribbon will deform when subject to  arbitrary surface stress and/or internal residual strains.  It establishes a quantitative relationship between surface stress, residual strain, elasticity, and helical properties such as helix angle, radius, axis, and chirality. The morphological transition from cylindrical helical ribbons to purely twisted ribbons and tubular structures can be tuned by relevant geometric parameters. The predictions were validated experimentally with composites formed by bonding a pre-stretched sheet (or two sheets) of elastomer to an elastic layer of soft acrylic.  The current results establish a new theoretical framework for predicting and tuning the shape of stress-driven helical structures for use in a broad spectrum of biological and engineering applications.

\section{Appendix: Mathematical description of the ribbon surface}
\label{Sec_surface}
A mapping of an initially straight ribbon into the as-deformed, helical one, can be constructed as follows.  First, recall that the centerline coordinates are given in Eq.~(\ref{eq:kinematics}).  Now, consider a material point along the short edge of the ribbon in the undeformed configuration given by $\textbf{Q}^0(s=0, t) = \textbf{P}(0) + t\textbf{B}(0)$.  Repeating the derivation in Eqs.~(\ref{curvatures}) through (\ref{eq:kinematics}) in terms of $t$, where $-w/2 \leq t \leq w/2$, yields the following expression for the edge space curve upon deformation:
\begin{equation}
\textbf{Q}(s=0,t) = \tilde{X}_0(t) \Ex + \tilde{Y}_0(t) \Ey +  \tilde{Z}_0(t) \Ez ,
\label{Q_Sol}
\end{equation}
where
\begin{eqnarray}
\tilde{X}_0(t) &=&  \frac{\beta}{\alpha^3} (\alpha t - \sin \alpha t) (\kappa_1 - \kappa_2)\sin \phi \cos \phi\nonumber\\
\tilde{Y}_0(t) &=&  t - \frac{\beta^2}{\alpha^3} (\alpha t - \sin \alpha t)  \nonumber\\
\tilde{Z}_0(t) &=&  \frac{\beta}{\alpha^2} (\cos \alpha t - 1) \,,
\label{eq:tkinematics}
\end{eqnarray}
with $\alpha$ and $\beta$ given in Eqs.~(\ref{eq:alpha}) and (\ref{eq:beta}), respectively. Now, the ribbon surface can be constructed by ``attaching'' the space curve described by Eqs.~(\ref{Q_Sol}) and ({\ref{eq:tkinematics}) to each point $s$ along the ribbon centerline, and locally rotating the space curve as dictated by the Frenet-Serret frame $\{ \textbf{T}(s), \textbf{N}(s), \textbf{B}(s) \}$.   That is, material points ${\bf Q}^0(s, t)$ in the undeformed configuration are mapped to ${\bf Q}(s, t) = X(s,t) \Ex + Y(s,t) \Ey + Z(s,t) \Ez$ upon deformation, such that
\begin{eqnarray}
\textbf{Q}(s,t)  &= & \textbf{P}(s) + [\tilde{X}_0(t) \Ex + \tilde{Y}_0(t) \Ey +  \tilde{Z}_0(t) \Ez] \tilde{A}(s) \nonumber \\
 & = & X(s,t) \Ex + Y(s,t) \Ey +  Z(s,t) \Ez,
\label{Q_nSol}
\end{eqnarray}
where the rotation matrix $\tilde{A}(s)$ is given by
\begin{equation}
\begin{split}
\tilde{A}(s) =& T_x(s) \Ex \otimes \Ex + T_y(s) \Ex \otimes \Ey + T_z(s) \Ex \otimes \Ez\\
&+ B_y(s) \Ey \otimes \Ex + B_y(s) \Ey \otimes \Ey + B_z(s) \Ey \otimes \Ez\\
&+ N_x(s) \Ez \otimes \Ex + N_y(s) \Ez \otimes \Ey + N_z(s) \Ez \otimes \Ez\\
=& \left[1 - \frac{\beta^2}{\alpha^2}(1 - \cos \alpha s)\right] \Ex \otimes \Ex + \frac{\beta \tau}{\alpha^2}(1 - \cos \alpha s) \Ex \otimes \Ey - \frac{\beta}{\alpha} \sin \alpha s \Ex \otimes \Ez\\
&+ \frac{\beta \tau}{\alpha^2}(1 - \cos \alpha s) \Ey \otimes \Ex + \left[\frac{\beta^2}{\alpha^2}(1 - \cos \alpha s) + \cos \alpha s \right] \Ey \otimes \Ey + \frac{\tau}{\alpha} \sin \alpha s \Ey \otimes \Ez\\
&+ \frac{\beta}{\alpha} \sin \alpha s \Ez \otimes \Ex - \frac{\tau}{\alpha} \sin \alpha s \Ez \otimes \Ey + \cos \alpha s \Ez \otimes \Ez.
\end{split}
\end{equation}
Finally, carrying out the matrix multiplications yields the following explicit expressions for $X(s,t)$, $Y(s,t)$ and $Z(s,t)$:
\begin{eqnarray}
X(s,t) &=& X_0(s) + \frac{\beta \tau}{\alpha^3}(\alpha t - \sin \alpha t) \left[1 + \frac{2 \beta^2}{\alpha^2}(\cos \alpha s - 1) \right]  \nonumber \\
&+& \frac{\beta}{\alpha} \left[(1 - \cos \alpha s)\tau t - \frac{\beta}{\alpha} (1 - \cos \alpha t) \sin \alpha s \right], \nonumber\\
Y(s,t) &=& Y_0(s)  + \frac{\beta^2}{\alpha^5}[(\tau^2 - \beta^2)(1 - \cos \alpha s) - \alpha^2  \cos \alpha s] (\alpha t - \sin \alpha t) \nonumber \\
&+& t \left[\frac{\beta^2}{\alpha^2}(1 - \cos \alpha s)  + \cos \alpha s \right]  +  \frac{\beta \tau}{\alpha^3}(1 - \cos \alpha t) \sin \alpha s  \nonumber \\
Z(s,t) &=& Z_0(s) +  \frac{\beta}{\alpha^2}  (\cos \alpha t - 1) \cos \alpha s \nonumber \\
&+& \frac{\tau}{\alpha^4}  [-2\beta^2 (\alpha t - \sin \alpha t)  + \alpha^3 t] \sin \alpha s \,,
\label{eq:Surface}
\end{eqnarray}
where $X_0(s)$, $Y_0(s)$, and $Z_0(s)$ are given in Eq.~(\ref{eq:kinematics}).

\appendix{Mathematical description of the ribbon surface}
\label{Sec_surface}
A mapping of an initially straight ribbon into the as-deformed, helical one, can be constructed as follows.  First, recall that the centerline coordinates are given in Eq.~(\ref{eq:kinematics}).  Now, consider a material point along the short edge of the ribbon in the undeformed configuration given by $\textbf{Q}^0(s=0, t) = \textbf{P}(0) + t\textbf{B}(0)$.  Repeating the derivation in Eqs.~(\ref{curvatures}) through (\ref{eq:kinematics}) in terms of $t$, where $-w/2 \leq t \leq w/2$, yields the following expression for the edge space curve upon deformation:
\begin{equation}
\textbf{Q}(s=0,t) = \tilde{X}_0(t) \Ex + \tilde{Y}_0(t) \Ey +  \tilde{Z}_0(t) \Ez ,
\label{Q_Sol}
\end{equation}
where
\begin{eqnarray}
\tilde{X}_0(t) &=&  \frac{\beta}{\alpha^3} (\alpha t - \sin \alpha t) (\kappa_1 - \kappa_2)\sin \phi \cos \phi\nonumber\\
\tilde{Y}_0(t) &=&  t - \frac{\beta^2}{\alpha^3} (\alpha t - \sin \alpha t)  \nonumber\\
\tilde{Z}_0(t) &=&  \frac{\beta}{\alpha^2} (\cos \alpha t - 1) \,,
\label{eq:tkinematics}
\end{eqnarray}
with $\alpha$ and $\beta$ given in Eqs.~(\ref{eq:alpha}) and (\ref{eq:beta}), respectively. Now, the ribbon surface can be constructed by ``attaching'' the space curve described by Eqs.~(\ref{Q_Sol}) and ({\ref{eq:tkinematics}) to each point $s$ along the ribbon centerline, and locally rotating the space curve as dictated by the Frenet-Serret frame $\{ \textbf{T}(s), \textbf{N}(s), \textbf{B}(s) \}$.   That is, material points ${\bf Q}^0(s, t)$ in the undeformed configuration are mapped to ${\bf Q}(s, t) = X(s,t) \Ex + Y(s,t) \Ey + Z(s,t) \Ez$ upon deformation, such that
\begin{eqnarray}
\textbf{Q}(s,t)  &= & \textbf{P}(s) + [\tilde{X}_0(t) \Ex + \tilde{Y}_0(t) \Ey +  \tilde{Z}_0(t) \Ez] \tilde{A}(s) \nonumber \\
 & = & X(s,t) \Ex + Y(s,t) \Ey +  Z(s,t) \Ez,
\label{Q_nSol}
\end{eqnarray}
where the rotation matrix $\tilde{A}(s)$ is given by
\begin{equation}
\begin{split}
\tilde{A}(s) =& T_x(s) \Ex \otimes \Ex + T_y(s) \Ex \otimes \Ey + T_z(s) \Ex \otimes \Ez\\
&+ B_y(s) \Ey \otimes \Ex + B_y(s) \Ey \otimes \Ey + B_z(s) \Ey \otimes \Ez\\
&+ N_x(s) \Ez \otimes \Ex + N_y(s) \Ez \otimes \Ey + N_z(s) \Ez \otimes \Ez\\
=& \left[1 - \frac{\beta^2}{\alpha^2}(1 - \cos \alpha s)\right] \Ex \otimes \Ex + \frac{\beta \tau}{\alpha^2}(1 - \cos \alpha s) \Ex \otimes \Ey - \frac{\beta}{\alpha} \sin \alpha s \Ex \otimes \Ez\\
&+ \frac{\beta \tau}{\alpha^2}(1 - \cos \alpha s) \Ey \otimes \Ex + \left[\frac{\beta^2}{\alpha^2}(1 - \cos \alpha s) + \cos \alpha s \right] \Ey \otimes \Ey + \frac{\tau}{\alpha} \sin \alpha s \Ey \otimes \Ez\\
&+ \frac{\beta}{\alpha} \sin \alpha s \Ez \otimes \Ex - \frac{\tau}{\alpha} \sin \alpha s \Ez \otimes \Ey + \cos \alpha s \Ez \otimes \Ez.
\end{split}
\end{equation}
Finally, carrying out the matrix multiplications yields the following explicit expressions for $X(s,t)$, $Y(s,t)$ and $Z(s,t)$:
\begin{eqnarray}
X(s,t) &=& X_0(s) + \frac{\beta \tau}{\alpha^3}(\alpha t - \sin \alpha t) \left[1 + \frac{2 \beta^2}{\alpha^2}(\cos \alpha s - 1) \right]  \nonumber \\
&+& \frac{\beta}{\alpha} \left[(1 - \cos \alpha s)\tau t - \frac{\beta}{\alpha} (1 - \cos \alpha t) \sin \alpha s \right], \nonumber\\
Y(s,t) &=& Y_0(s)  + \frac{\beta^2}{\alpha^5}[(\tau^2 - \beta^2)(1 - \cos \alpha s) - \alpha^2  \cos \alpha s] (\alpha t - \sin \alpha t) \nonumber \\
&+& t \left[\frac{\beta^2}{\alpha^2}(1 - \cos \alpha s)  + \cos \alpha s \right]  +  \frac{\beta \tau}{\alpha^3}(1 - \cos \alpha t) \sin \alpha s  \nonumber \\
Z(s,t) &=& Z_0(s) +  \frac{\beta}{\alpha^2}  (\cos \alpha t - 1) \cos \alpha s \nonumber \\
&+& \frac{\tau}{\alpha^4}  [-2\beta^2 (\alpha t - \sin \alpha t)  + \alpha^3 t] \sin \alpha s \,,
\label{eq:Surface}
\end{eqnarray}
where $X_0(s)$, $Y_0(s)$, and $Z_0(s)$ are given in Eq.~(\ref{eq:kinematics}).

{\it Acknowledgements} -- The authors would like to thank Q. Guo, Y. Yu and W. Shan, for their assistance in the experimental demonstration, and C. Li, for useful discussions.  This work has been supported, in part, by the Princeton Institute for the Science and Technology of Materials (PRISM) at Princeton University, the Sigma Xi Grants-in-Aid of Research (GIAR) program, and American Academy of Mechanics Founder's Award from the Robert M. and Mary Haythornthwaite Foundation. Z. Chen is supported by a Society in Science - Branco Weiss fellowship.

\label{lastpage}
\end{document}